\documentstyle[12pt]{article}
\newcommand{\alt}{\mathrel{\raisebox{-.6ex}{$\stackrel{\textstyle<}{\sim}$}}}

\catcode`@=11
\newcount\@tempcntc
\def\@citex[#1]#2{\if@filesw\immediate\write\@auxout{\string\citation{#2}}\fi
  \@tempcnta\z@\@tempcntb\m@ne\def\@citea{}\@cite{\@for\@citeb:=#2\do
    {\@ifundefined
       {b@\@citeb}{\@citeo\@tempcntb\m@ne\@citea\def\@citea{,}{\bf ?}\@warning
       {Citation `\@citeb' on page \thepage \space undefined}}%
    {\setbox\z@\hbox{\global\@tempcntc0\csname b@\@citeb\endcsname\relax}%
     \ifnum\@tempcntc=\z@ \@citeo\@tempcntb\m@ne
       \@citea\def\@citea{,}\hbox{\csname b@\@citeb\endcsname}%
     \else
      \advance\@tempcntb\@ne
      \ifnum\@tempcntb=\@tempcntc
      \else\advance\@tempcntb\m@ne\@citeo
      \@tempcnta\@tempcntc\@tempcntb\@tempcntc\fi\fi}}\@citeo}{#1}}
\def\@citeo{\ifnum\@tempcnta>\@tempcntb\else\@citea\def\@citea{,}%
  \ifnum\@tempcnta=\@tempcntb\the\@tempcnta\else
   {\advance\@tempcnta\@ne\ifnum\@tempcnta=\@tempcntb \else \def\@citea{--}\fi
    \advance\@tempcnta\m@ne\the\@tempcnta\@citea\the\@tempcntb}\fi\fi}
\catcode`@=12

\begin{document}

\font\fortssbx=cmssbx10 scaled \magstep2
\hbox to \hsize{
\includegraphics{uwlogo.ps}
\hskip.5in \raise.1in\hbox{\fortssbx University of Wisconsin - Madison}
\hfill$\vcenter{\hbox{\bf MADPH-96-936}
                \hbox{\bf UPR-0696T}
                \hbox{\bf UTEXAS-HEP-96-2}
\hbox{\bf DOE-ER-40757-078}
                \hbox{April 1996}}$ }

\vspace*{.5in}

\begin{center}
{\large
Baryonic $Z'$ connection of LEP $R_{b,c}$ data with Tevatron $(W,Z,\gamma)b\bar
b$ events}\\[5mm]
V. Barger$^{a)}$, Kingman Cheung$^{b)}$, and Paul Langacker$^{c)}$\\[3mm]
\it
$^{a)}$Dept. of Physics, University of Wisconsin, Madison, WI 53706\\
$^{b)}$Center for Particle Physics, University of Texas, Austin, TX 78712\\
$^{c)}$Dept. of Physics and Astronomy, University of Pennsylvania, Philadelphia
PA 19104
\end{center}

\vspace{1in}

\begin{abstract}
The mixing of a new $Z'$ boson with the $Z$ significantly improves the fit to
the LEP precision electroweak data, provided that the $Z'$ couples mainly to
quarks. If $M_{Z_2}<200$~GeV, the $s$-channel $Z_2$ production and
$(W,Z,\gamma)Z_2$ pair production cross sections
at the Tevatron give an excess above QCD of $b\bar b$ and $(W,Z,\gamma)b\bar b$
events, respectively,  with invariant mass $m(b\bar b)\approx M_{Z_2}$, which
provide viable signals for detection of the $Z_2$. The interference of the
$Z_2$ with $\gamma,Z_1$ in $e^+e^-\to \bar bb(\bar cc)$ at LEP\,1.5 energies is
correlated with $R_b(R_c)$ and may be observable.
\end{abstract}

\thispagestyle{empty}
\newpage

The Standard Model (SM) has long provided an excellent representation of
particle interactions. Recently, however, possible indications of discrepancies
with SM predictions have surfaced in LEP\cite{lep} data. The
LEP measurements\cite{lep} of
\begin{equation}
R_{b(c)} = \Gamma(Z\to \bar bb(\bar cc)) / \Gamma(Z\to \rm hadrons)
\end{equation}
deviate by $3.7\sigma\,(-2.4\sigma)$ from the
SM\cite{ew-lang,ew-hagi}.\footnote{The values in the Spring 1996 preliminary
update\cite{lep-ew} are slightly closer to the SM, deviating by $3.5\sigma\,
(-1.8\sigma)$.} These deviations have generated a flurry of phenomenological
activity
since they may be the first indications of physics beyond the SM. Proposed
explanations of the observed phenomena include supersymmetric\cite{susy} or
other new particles\cite{newparticles},  extra $Z$
bosons\cite{bcl,nonunivz,chiap,alta,babu,carone,malwaki,zprime,frampton,agash},
technicolor\cite{techni}, and other\cite{misc} models.
Our interest here is in possible extra $Z$ boson interpretations, which have
immediate implications for physics at the Tevatron.
We point out that $s$-channel $Z_2$ production and the pair production
processes $(W,Z,\gamma)Z_2$ with $Z_2\to b\bar b$ decays will lead to $b\bar b$
and $(W,Z,\gamma)b\bar b$ events at the Tevatron, with a $b\bar b$ invariant
mass peaked at $M_{Z_2}$ in excess of QCD backgrounds if $M_{Z_2}\alt 200$~GeV.
Here we use $Z_2$ to denote the mass eigenstate of the heavy $Z$ boson
after $Z-Z'$ mixing. $Z_2$ interference effects may be observable in $e^+e^-\to \bar bb(\bar cc)$ at LEP\,1.5 energies.

Our work  has a distinct vantage point from other recent $Z'$ analyses of the
$R_{b,c}$ data that advocate a $Z'$ boson with mass $\approx
1$~TeV\cite{chiap,alta} to account
for an excess above QCD of the inclusive jet cross section at $E_T>200$~GeV
reported by CDF\cite{cdf}. Although quark distributions are well constrained by
deep inelastic scattering data\cite{glover}, a smooth rise in the $E_T$ jet
cross section compared to QCD expectations can possibly be explained by other
means, such as a modification of the gluon structure function at high
$x$\cite{cteq-high-et} or a flattening of $\alpha_s(Q^2)$ at high $Q^2$ due to
new particles\cite{vb+white}. Also the CDF high $E_T$ jet anomaly is not
present in preliminary D0 data\cite{dzero}.

A large class of string models with supersymmetry contain additional $U(1)'$
symmetries and additional exotic matter multiplets. In many of these models the
$Z'$ and exotic masses are either of ${\cal O}(M_Z)$ or of order $10^{8}$ to
$10^{14}$~GeV\cite{cvet-lang}. Consequently a search for $Z'$ bosons in the
electroweak mass region $\alt 1$~TeV is well motivated.
Through the mixing of the $Z'$ boson with the $Z$, the predictions of
electroweak observables are
modified\cite{rosner-z',barger-z',durkin,hew-rizz}. Thus, it is natural to
see if  $Z,Z'$ mixing effects can better account for the precision electroweak
measurements. In general, this mixing affects both lepton and quark partial
widths of the $Z$ as well as the total width. The changes in the widths vary
from model to model because of the different chiral couplings. In the usual
models based on grand unification with $SO(10)$ or $E_6$ gauge groups (without
kinetic mixing\cite{holdom}), all $Z$
partial widths are modified. However, because the leptonic widths agree well
with SM predictions, an overall fit to the electroweak data is then not
significantly improved by $Z'$ mixing in these models and the $R_b$ excess is
not explained.

If, however, we consider a model in which the $Z'$ couples solely or dominantly
to quarks with a universal strength, a
substantial improvement results in the description of the precision electroweak
data, as detailed below, and found in other recent
analyses\cite{bcl,chiap,alta,babu,agash}.
A reasonable fit to the data is
obtained for a range of universal chiral couplings of the $Z'$ boson.  A gauge
symmetry generated by baryon number, $U(1)_B$, is an interesting
possibility\cite{carone}, since this avoids potential problems associated with
the breaking of global baryon number by quantum gravity effects (e.g., an
unacceptable proton decay rate in supersymmetric theories). In this case the
$Z'$ has vector couplings. Another possibility is kinetic mixing of the two
U(1)'s\cite{babu,holdom} to suppress the leptonic couplings.
The $U(1)_\eta$ model of $E_6$ is an interesting model in which this may
occur\cite{babu}. Here the cancellation of contributions to the
$Z$-leptonic width is fine tuned and leptonic $Z_2$ decays may still be present
at a suppressed level. In the following, we will consider family-universal
couplings to baryon and axial-baryon number. As in
Refs.~\cite{chiap,alta}, we assume that the model can be embedded in an
anomaly-free theory. Extension of the results to models (such as $U(1)_\eta$)
with different couplings to charge 2/3 and $-1/3$ quarks, or to the family
non-universal case, is straightforward.

A $Z'$ coupled to quarks has very interesting implications for physics at the
Tevatron collider. If its mass is $\alt 200$~GeV, it could be
produced in the $s$-channel and in conjunction with the $W,\ Z$ or $\gamma$ and
detected via its
$Z_2\to b\bar b$ decay mode (and possibly also through $Z_2\to c\bar c$). The
signatures for $WZ_2$ and $ZZ_2$ production
would be similar to Higgs boson production $WH$ and $ZH$, with $H\to b\bar b$
decays, but the $Z_2$ signals could be considerably higher.

\section*{\bf $Z-Z'$ mixing}

Following the notation of Ref.~\cite{lang-luo}, the Lagrangian describing the
neutral current gauge interactions of the
standard electroweak SU(2)$\times$ U(1) and extra U(1)'s is given by
\begin{equation}
\label{L}
- {\cal L}_{\rm NC} = e J_{\rm em}^\mu A_\mu + \sum_{\alpha=1}^{n}
g_\alpha J^\mu_\alpha Z^0_{\alpha \mu}\;,
\end{equation}
where $Z^0_1$ is the SM $Z$ boson and $Z^0_\alpha$ with $\alpha\ge 2$ are the
extra $Z$ bosons in the weak-eigenstate basis\cite{reviews}.
In our case, we only
consider one extra $Z_2^0$ mixing with the SM $Z^0_1$ boson.  The coupling
constant $g_1$ is the SM coupling $g/\cos\theta_{\rm w}$. For grand unified
theories (GUT) $g_2$ is related to $g_1$ by
\begin{equation}
\frac{g_2}{g_1} = \left(\frac{5}{3}\, x_{\rm w} \lambda\right)^{1/2} \simeq
0.62\lambda^{1/2} \,,
\label{eq:g2/g1}
\end{equation}
where $x_{\rm w}=\sin^2\theta_{\rm w}$ and $\theta_{\rm w}$ is the weak mixing
angle.
The factor $\lambda$ depends on the symmetry breaking pattern and the fermion
sector of the theory but is usually of order unity.

Since we only consider the mixing of $Z_1^0$ and $Z_2^0$ we can rewrite
the Lagrangian in Eq.~(\ref{L}) with only the $Z^0_1$ and $Z^0_2$
interactions
\begin{equation}
\label{ZZ}\textstyle
-{\cal L}_{Z^0_1 Z^0_2} = g_1 Z^0_{1\mu} \left[ \frac{1}{2} \sum_i
 \bar \psi_i \gamma^\mu (g_v^{i(1)} - g_a^{i(1)} \gamma^5 ) \psi_i \right] +
 g_2 Z^0_{2\mu} \left[ \frac{1}{2} \sum_i
 \bar \psi_i \gamma^\mu (g_v^{i(2)} - g_a^{i(2)} \gamma^5 ) \psi_i \right]\;,
\end{equation}
where for both quarks and leptons
\begin{equation}
g_v^{i(1)} = T_{3L}^i - 2 x_{\rm w} Q_i \,, \qquad
g_a^{i(1)} = T_{3L}^i\,,
\end{equation}
and we consider the case in which $Z_2$ couples only to quarks,
\begin{equation}
g_v^{q(2)} = \epsilon_V \,,\qquad
g_a^{q(2)} = \epsilon_A \,,\qquad
g_v^{\ell(2)} = g_a^{\ell(2)} = 0\;.
\end{equation}
Here $T_{3L}^i$ and $Q_i$ are, respectively, the third component of the weak
isospin and the electric charge of the fermion $i$; $\epsilon_V$ and
$\epsilon_A$ are parameters of the $Z_2$ sector.
The mixing of the weak eigenstates $Z^0_1$ and $Z^0_2$ to form  mass eigenstates $Z_1$ and $Z_2$  can be parametrized
by a mixing angle $\theta$
\begin{equation}
\label{mixing}
\left ( \begin{array}{c} Z_1 \\
                         Z_2
        \end{array} \right ) = \left( \begin{array}{cc}
                                  \cos\theta & \sin\theta \\
                                 -\sin\theta & \cos\theta
                                      \end{array} \right ) \;
    \left( \begin{array}{c} Z^0_1 \\
                            Z^0_2
            \end{array} \right ) \;.
\end{equation}
The mass of $Z_1$ is $M_{Z_1}=91.19$~GeV and $M_{Z_2}$ is unknown.

Substituting Eq.~(\ref{mixing}) into Eq.~(\ref{ZZ}) we obtain the interactions
of the mass eigenstates $Z_1$ and  $Z_2$ with fermions
\begin{equation}
\label{rule}
-{\cal L}_{Z_1 Z_2} = \sum_i \frac{g_1}{2} \biggl [
  Z_{1\mu} \bar \psi_i \gamma^\mu (v_s^i - a_s^i \gamma^5 ) \psi_i  +
  Z_{2\mu} \bar \psi_i \gamma^\mu (v_n^i - a_n^i \gamma^5 ) \psi_i  \biggr ]\,,
\end{equation}
where
\begin{eqnarray}
v_s^i = g_v^{i(1)} + \frac{g_2}{g_1} \, \theta \, g_v^{i(2)} \,, &&
a_s^i = g_a^{i(1)} + \frac{g_2}{g_1} \, \theta \, g_a^{i(2)} \,, \\
v_n^i = \frac{g_2}{g_1}\, g_v^{i(2)} - \theta \, g_v^{i(1)}\,, &&
a_n^i = \frac{g_2}{g_1}\, g_a^{i(2)} - \theta \, g_a^{i(1)}\,.
\end{eqnarray}
Here we use the valid approximation $\cos\theta\approx 1$ and $\sin\theta
\approx \theta$.
The Feynman rules for the interactions of $Z_1$ and $Z_2$ with the fermions
can be easily obtained from Eq.~(\ref{rule}).

\section*{\bf Precision Electroweak Constraints}

{}From studies of $Z'$ mixing effects on the $Z_1$ coupling, the products
$\theta\lambda^{1/2}\epsilon_V$ and $\theta\lambda^{1/2}\epsilon_A$ can be
determined. Without loss of generality we can take the $\epsilon_V$ and
$\epsilon_A$ to be normalized to unity and write
\begin{equation}
\epsilon_V = \sin\gamma\,,\qquad
\epsilon_A = \cos\gamma\,,\qquad
\kappa = -\theta \textstyle \left({5\over3}x_{\rm w}\lambda\right)^{1/2} \,.
\end{equation}
The partial widths for $Z_1$-decays to quarks are determined by the couplings
\begin{eqnarray}
v_s^b &=& \textstyle -{1\over2} + {2\over3}x_{\rm w} - \kappa \sin\gamma\ =\
-0.35 - \kappa\sin\gamma\,,\\
a_s^b &=& \textstyle - {1\over2} - \kappa \cos\gamma\ =\ -0.5 -
\kappa\cos\gamma \,,\\
v_s^c &=& \textstyle\phantom+ {1\over2} -{4\over3}x_{\rm w} - \kappa
\sin\gamma\ =\ \phantom+ 0.19 - \kappa \sin\gamma \,,\\
a_s^c &=& \textstyle\phantom+ {1\over2} - \kappa \cos\gamma\ =\ 0.5 -
\kappa\cos\gamma \,,
\end{eqnarray}
where the value $x_{\rm w}=0.23$ is used in the approximate equalities. The
modifications in the SM partial widths are then
\begin{eqnarray}
\delta\Gamma(Z\to b\bar b) &\simeq& \phantom+ \kappa C(M_{Z_1}^2)
\Gamma^0( 0.69\sin\gamma + 1.0\cos\gamma ) \,,\\
\delta\Gamma(Z\to c\bar c) &\simeq& -\kappa C(M_{Z_1}^2)
\Gamma^0( 0.39\sin\gamma + 1.0\cos\gamma )\,,
\end{eqnarray}
where
\begin{equation}
\Gamma^0 \equiv  {G_F M_{Z_1}^3 \over 2\sqrt2 \pi}\,
\end{equation}
and
\begin{equation}
C(Q^2) = 1 + {\alpha_s\over\pi} + 1.409{\alpha_s^2\over\pi^2} -
12.77{\alpha_s^3\over\pi^3} \,,
\end{equation}
with $\alpha_s = \alpha_s(Q^2)$\cite{ew-lang}.
Fermion mass corrections, effects related to the shift induced in $M_Z$ by the
mixing\cite{lang-luo}, and electroweak corrections are not displayed for
simplicity but are incorporated in the numerical analysis.

Similar results apply for the other $T_3 = -1/2$ and 1/2 flavors, respectively.
The total hadronic width
\begin{equation}
\Gamma(Z\to{\rm hadrons}) = 3\Gamma(Z\to b\bar b) + 2\Gamma(Z\to c\bar c)
\end{equation}
is modified by
\begin{equation}
\delta\Gamma_{\rm had} = \kappa C \Gamma^0 (1.3\sin\gamma + 1.0\cos\gamma) \,.
\end{equation}
Thus, a vector baryonic $Z'$ $(\lambda=\pi/2)$ gives the modifications
\begin{equation}
\delta\Gamma(Z\to b\bar b) \simeq  0.69 \kappa C \Gamma^0 \,,\quad
\delta\Gamma(Z\to c\bar c) \simeq -0.39 \kappa C \Gamma^0 \,,\quad
\delta\Gamma_{\rm had} \simeq  1.3 \kappa C \Gamma^0 \,.
\end{equation}
The $Z\to b\bar b$ partial width is increased (for $\kappa>0$) and the $Z\to
c\bar c$ partial
width is decreased, which are the directions of the deviations from SM
predictions indicated by the LEP data. The increase in the total hadronic width
can be compensated by  a smaller value of $\alpha_s(M_Z^2)$ in $C(M_{Z_1}^2)$
than that obtained in the SM fits;
then both $\Gamma_{\rm had}$ and $\Gamma_{\rm tot}$ measurements are well
described by the $Z'$ mixing model.

An axial baryonic $Z'$ ($\gamma=0$) gives the changes
\begin{equation}
\delta\Gamma(Z\to b\bar b) \simeq  1.0\kappa C\Gamma^0\,,\quad
\delta\Gamma(Z\to c\bar c) \simeq -1.0\kappa C\Gamma^0\,,\quad
\delta\Gamma_{\rm had} \simeq  1.0\kappa C\Gamma^0\,.
\end{equation}
Here the effects in the $b\bar b$ and $c\bar c$ channels are again in the
desired direction (for $\kappa>0$), but larger, and the change in
$\delta\Gamma_{\rm had}$ is
somewhat less. A range of $\gamma$ values can produce fits that are
significantly better than the SM; we focus
on $\gamma=0$ and $\pi/2$ henceforth as representative cases. Even better fits
could be obtained by adjusting $\gamma$. We will also briefly consider the
fine-tuned case $\cot\gamma=-1.3$, for which the direct contribution to
$\delta\Gamma_{\rm had}$ vanishes.

We have made fits to the full set of electroweak measurements similar to
analyses of the SM\cite{ew-lang}. In particular, we include the (important)
constraints from deep inelastic neutrino scattering and atomic parity
violation, which were not included in the analyses of
Refs.~\cite{chiap,alta,babu}.
The best fit value of $\alpha_s(M_Z^2)$ comes out somewhat low for the pure
axial and pure vector cases, so we made
subsequent fits with $\alpha_s(M_Z^2)$ fixed at 0.11, 0.115, and 0.12. The
chi-square for the axial model is moderately increased by fixing
$\alpha_s$ at the higher values, while for the vector model the quality of the
fit decreases significantly. The values\footnote{The shifts in the $Z_1$
couplings depend only on the combination $\theta\lambda^{1/2}$. The shift in
$M_{Z_1}$ and the effects of $Z_2$ exchange have different dependences on
$\lambda$ and $\theta$. However, the global fit results are insensitive to
$\lambda$ in the range 0.25--4 except for the scaling of $\theta$ as
$\lambda^{-1/2}$. The fit results are given for $\lambda=1$.}
 for $\alpha_s,\ m_t,\ \theta\lambda^{1/2}$ and $M_2/M_1$ are listed in
Tables~\ref{table:gamma=0} and \ref{table:gamma=pi/2}.
Table~\ref{table:cotgamma} contains the results of the model with
$\cot\gamma=-1.3$.
Excellent fits are obtained with $\alpha_s$ close to the SM fit value
$\alpha_s=0.123$.
For comparison with the $\chi^2$ values in these fits, the
$\chi^2$ value found in the SM fit is $\chi^2=192$ for 208 degrees of freedom.
Table~\ref{table:compare}
compares the fit to the interesting observables $R_b,\ R_c,\ R_\ell,\
\Gamma_{\rm had},\ \Gamma_{\rm tot}$ and $A_{\rm FB}(b\bar b)$, where the
latter quantity is the $b\bar b$ asymmetry; the ``pull" of each of these
observables in the fit is given.

\begin{table}[h]
\centering
\caption{\label{table:gamma=0}
Parameters determined by electroweak data analysis for $\gamma=0$ ($Z'$
with pure axial vector coupling). The $\chi^2$ values are for 206 (207) degrees
of freedom for $\alpha_s$ free (fixed). The upper bounds on $M_2/M_1$ are one
sigma mass limits.}
\medskip
\tabcolsep=1.4em
\begin{tabular}{rllrcr}
\hline\hline
\multicolumn{1}{c}{$\sin^2\theta_{\rm w}$}& \multicolumn{1}{c}{$\alpha_s$}&
\multicolumn{1}{c}{$m_t$}& \multicolumn{1}{c}
{$\theta \lambda^{1/2}$} &
\multicolumn{1}{c}{$M_2/M_1$}& \multicolumn{1}{c}{$\chi^2$}\\
\hline
0.2313(2)& 0.095(8)&    $183^{+7}_{-11}$& $-$0.025(7)& $<$1.9&   176\\
0.2314(2)& 0.11 fixed&  $181^{+7}_{-10}$& $-$0.014(3)& $<$2.9&   179\\
0.2315(2)& 0.115 fixed& $180^{+7}_{-9}$&  $-$0.011(3)& $<$3.9&   182\\
0.2315(2)& 0.12 fixed&  $179^{+7}_{-9}$&  $-$0.007(3)& $<$6.3&   185\\
\hline\hline
\end{tabular}
\end{table}

\begin{table}[t]
\centering
\caption{\label{table:gamma=pi/2}
Parameters determined by electroweak data analysis for $\gamma=\pi/2$ ($Z'$
with pure  vector couplings).}
\medskip
\tabcolsep=1.4em
\begin{tabular}{rlllcr}
\hline\hline
\multicolumn{1}{c}{$\sin^2\theta_{\rm w}$}& \multicolumn{1}{c}{$\alpha_s$}&
\multicolumn{1}{c}{$m_t$}& \multicolumn{1}{c}
{$\theta \lambda^{1/2}$} &
\multicolumn{1}{c}{$M_2/M_1$}& \multicolumn{1}{c}{$\chi^2$}\\
\hline
0.2313(2)& 0.068(17)&   181(7)& $-$0.037(11)& $<$1.02&           179\\
0.2315(2)& 0.11 fixed&  179(7)& $-$0.010(2)&  $<$1.1\phantom0&   186\\
0.2315(2)& 0.115 fixed& 179(7)& $-$0.007(2)&  $<$1.15&           187\\
0.2315(2)& 0.12 fixed&  179(7)& $-$0.004(2)&  $<$1.4\phantom0&   189\\
\hline\hline
\end{tabular}
\end{table}

In the SM fit the Higgs mass is constrained to the range
\begin{equation}
60 < m_H < 100
\end{equation}
with the best fit at the lower end of the allowed range. This is driven mainly
by $R_b$ and the SLD polarization asymmetry\cite{ew-lang}. In the
mixing models the preference for any particular Higgs mass value is weakened
significantly, especially for the cases which come close to the experimental
$R_b$.
Our precision electroweak analysis does not include the case of an approximate
$Z_2,Z$ mass degeneracy\cite{ross}, since the LEP/SLD extractions of the
$Z$-parameters assume a single resonance description. Values of $M_{Z_2}<
M_{Z_1}$ area also possible, but we have not analyzed this case in detail.

\begin{table}[h]
\centering
\caption{\label{table:cotgamma}
Parameters determined for the model with no direct contribution to
$\delta\Gamma_{\rm had}$ ($\cot\gamma=-1.3$). The standard model fit
parameters are also shown. The  $\chi^2$ are for 206 (208) degrees of freedom.}
\tabcolsep=1em
\medskip
\begin{tabular}{lcccccc}
\hline\hline
& $\sin^2\theta_{\rm w}$& $\alpha_s$& $m_t$& $\theta\lambda^{1/2}$&
$M_2/M_1$& $\chi^2$\\
\hline
$\delta\Gamma_{\rm had}=0$& 0.2312(2)& 0.121(4)& $185^{+7}_{-8}$& $-0.043(11)$&
$<1.2$& 177\\
SM& 0.2315(2)& 0.123(4)& 180(7)& ---& ---& 192\\
\hline\hline
\end{tabular}
\end{table}

\begin{table}[t]
\centering
\caption{\label{table:compare}
Comparison of fits to the observables $R_b,\ R_c,\ R_\ell,\ \Gamma_{\rm had},\
\Gamma_{\rm tot}$ and $A_{\rm FB}(b\bar b)$. ($\Gamma_{\rm had}$ is actually a
quantity derived from the standard fit variables\protect\cite{lep,ew-lang}.)
For the vector and axial cases, the results are for $\alpha_s = 0.110$.
Widths are in GeV.
For each fit the ``pull", i.e., (fit value $-$ expt.\ value)/error, is shown in
square brackets.}
\medskip
\tabcolsep=.7em
\begin{tabular}{llllll}
\hline\hline
& \multicolumn{5}{c}{Baryonic $Z'$ Model}\\
& \multicolumn{1}{c}{Expt.}& \multicolumn{1}{c}{SM}&
\multicolumn{1}{c}{$\gamma=0$}& \multicolumn{1}{c}{$\gamma=\pi/2$}&
\multicolumn{1}{c}{$\cot\gamma=-1.3$}\\
\noalign{\vskip-1ex}
& \multicolumn{1}{c}{(LEP+SLD)}& &\multicolumn{1}{c}{(axial)}&
\multicolumn{1}{c}{(vector)}&
\multicolumn{1}{c}{($\delta\Gamma_{\rm had}=0$)}\\
\hline
$R_b$& 0.2219(17)& 0.2155 [$-3.8$]& 0.2194 [$-1.5$]& 0.2170 [$-2.9$]& 0.2210
[$-0.5$]\\
$R_c$& 0.1540(74)& 0.172 [2.4]& 0.166 [1.6]& 0.170 [2.2]& 0.164 [1.4]\\
$R_\ell$& 20.788(32)& 20.77 [$-0.6$]& 20.79 [0.1]& 20.78 [$-0.3$]& 20.78
[$-0.3]$\\
$\Gamma_{\rm had}$& 1.7448(30)& 1.746 [0.4]& 1.747 [0.7]& 1.747 [0.7]& 1.744
[$-0.3$]\\
$\Gamma_{\rm tot}$& 2.4963(32)& 2.500 [1.2]& 2.501 [1.5]& 2.500 [1.2]& 2.497
[0.2]\\
$A_{\rm FB}(b\bar b)$& 0.0997(31)& 0.102 [0.7]& 0.102 [0.7]& 0.102 [0.7]& 0.100
[0.1]\\
\hline\hline
\end{tabular}
\end{table}

\section*{\bf $Z_2$ Decays}

The decay width of
$Z_2 (Z_1) \to f \bar f$ is given by
\begin{equation}
\label{width}
\Gamma (Z_{2(1)} \to f\bar f ) = \frac{G_F M_{Z_1^0}^2 }{6\pi \sqrt{2} }
N_c  C(M_{Z_{2(1)}}^2) M_{Z_{2(1)}} \sqrt{ 1 - 4x} \left[
v_{n(s)}^{f2} (1+2x) +  a_{n(s)}^{f2} (1-4x) \right] \,,
\end{equation}
where $x=m_f^2/M_{Z_{2(1)}}^2$, $N_c=3$ or 1 if $f$ is a quark or a lepton,
respectively, $G_F$ is the Fermi coupling constant, and $M_{Z_1^0}$ is the SM
$Z$ mass.
We calculated $\alpha_s(M_{Z_2})$ from the two-loop expression with
$\Lambda_{\rm QCD}=200$~MeV and 5 flavors for $M_{Z_2}<2m_t$ and 6 flavors
above $2m_t$.

The $Z_2$ width is proportional to $\lambda$, which sets the strength of
the $Z_2$ coupling; see Eq.~(\ref{eq:g2/g1}). For $\lambda=1$ the total $Z_2$
width is
\begin{equation}
\Gamma_{Z_2} / M_{Z_2} = 0.022 \quad {\rm for}\ M_{Z_2} < 2m_t \,,\qquad
\Gamma_{Z_2} / M_{Z_2} = 0.026 \quad {\rm for}\ M_{Z_2} > 2m_t \,.
\end{equation}
The widths would be increased somewhat if there are open channels for decay
into superpartners or exotic particles.

\section*{\bf $Z_2$ Production in the $s$-channel}

The $Z_2$ state can be directly produced at a hadron collider via the $q\bar
q\to Z_2$ subprocesses, for which the cross section in the narrow $Z_2$ width
approximation is\cite{collider}
\begin{equation}
\hat\sigma(q\bar q\to Z_2) = K {2\pi\over3} {G_F\,M_{Z_1}^2\over\sqrt2}
\left[ \left(v_n^q\right)^2 + \left(a_n^q\right)^2 \right]
 \delta\! \left(\hat s - M_{Z_2}^2\right) \,.
\end{equation}
The $K$-factor represents the enhancement from higher order QCD processes,
estimated to be\cite{collider}
%
$K = 1 + {\alpha_s(M_{Z_2}^2)\over2\pi} \textstyle {4\over3} \left( 1 +
{4\over3}\pi^2 \right) \simeq 1.3$.
%
In the approximation that the terms proportional to $\theta$ in the couplings
$v_n^q,\ a_n^q$ are neglected,
\begin{equation}
\left(v_n^q\right)^2 + \left(a_n^q\right)^2 = (0.62)^2 \lambda
\end{equation}
and the cross section is independent of the parameter $\gamma$.

The jet-jet invariant mass resolution smearing of hadron collider detectors is
typically $\Delta m(jj)/m(jj)=0.1$,
which includes the effects of  QCD radiation and detector smearing.
Since this mass resolution well exceeds the $Z_2$
width when the $Z_2$ is ${\cal O}(M_Z)$,  we include a Gaussian
smearing of $m(jj)$ with this rms resolution in calculating $m(jj)$
distributions associated with $Z_2$ decays.

In calculating the QCD background to $s$-channel $Z_2$ production we
include
interference effects and  calculate the $q\bar q\to q\bar q$ process at
the amplitude level, including $Z_2,\ Z$ and $\gamma$ exchanges along with the
QCD gluon exchange amplitudes. The non-interfering backgrounds from the
$W$-exchange processes
$q\bar q'\to q\bar q'$ and $q\bar q'\to q'\bar q$ and the backgrounds from
$gg,\ gq$, and $g\bar q$ initiated processes are added to get the full
jet-jet cross section. In our calculations we use the CTEQ3L parton
distributions of Ref.~\cite{cteq}.

The UA2 Collaboration\cite{ua2} has detected the $W+Z$ signal in the dijet mass
region $48<m(jj)<138$~GeV and has placed upper bounds on $\sigma B(Z_2\to jj)$
over the range $80<m(jj)<320$~GeV. Figure~\ref{fig:Z2_UA2} compares our $Z_2$
model predictions for $\lambda=0.2,\ 0.6,$ and 1 with the UA2 upper bounds. We
see that a $\lambda$ upper bound of order 0.7 to 1 is indicated for $100<M_{Z_2}<180$~GeV.
However, because of the uncertainty in the $K$-factor in the theoretical cross
section calculation and the difficulty in obtaining an experimental bound by
subtraction of a smooth background, we subsequently consider $\lambda=1$  at
any $M_{Z_2}$ for illustration.

Inclusive $Z_2$ production with $Z_2\to b \bar b$ decays may be detectable at
the Tevatron as an excess of events in the $b\bar b$ invariant mass
distribution at $m(b\bar b)\approx M_{Z_2}$ and in the inclusive transverse
momentum distribution of the $b$, which has a Jacobian peak at
$p_T(b)\simeq{1\over2}M_{Z_2}$. These distributions are  illustrated for
leading order QCD in Fig.~\ref{fig:Z2->bbar}. Vertex and semileptonic tagging
of the $b$'s can be used to reject the backgrounds from other quarks and
gluons. The backgrounds due to $gg\to b\bar b$ production are nonetheless very
large, so identification of the signal contribution here is difficult.

The $Z_2$ can be produced in $e^+e^-$ collisions via any direct $e^+e^-$
coupling and its $e^+e^-$ coupling induced by mixing. Here we consider the
$e^+e^-$ coupling that results solely from mixing. Figure~\ref{fig:resonance}
illustrates the effects of a $M_{Z_2}=105$~GeV resonance on the $e^+e^-\to
b\bar b$ and $c\bar c$ cross sections and on the $e^+e^-\to \bar bb$
forward-backward asymmetry $A_{\rm FB}$. An interference of $Z_2,\ Z_1$ and
$\gamma$ contributions gives the wave-like structure, with the interference
vanishing close to $\sqrt s = M_{Z_2}$. The signs of the interference
contributions are opposite in $b\bar b$ and $c\bar c$ (and are related at
$\sqrt s < M_{Z_2}$ to the signs of the deviations of $R_b$ and $R_c$ from the
SM). Consequently, flavor identification is necessary to observe the effect.
To quantify the effect, we take the difference of cross-sections at $\pm 1$~GeV on either side of the interference zero. In the $M_{Z_2}=105$~GeV illustration the values of 
$\Delta \sigma = \sigma(\sqrt{s}=104\,{\rm GeV}) - \sigma(\sqrt{s}=106 \,
{\rm GeV})$ are
\begin{equation}
\begin{array}{ccc} 
\Delta \sigma^{SM+Z_2(axial)}_{b\bar b} = 39\;{\rm pb}; &
\Delta \sigma^{SM+Z_2(vector)}_{b\bar b} = 34\;{\rm pb}; &
\Delta \sigma^{SM}_{b\bar b} = 22\;{\rm pb}; \\ 
\Delta \sigma^{SM+Z_2(axial)}_{c\bar c} = 0.72\;{\rm pb}; &
\Delta \sigma^{SM+Z_2(vector)}_{c\bar c} = 11\;{\rm pb}; &
\Delta \sigma^{SM}_{c\bar c} = 18\;{\rm pb}. 
\end{array}
\end{equation}
 A fine energy scan at LEP\,1.5 in the region of $\sqrt s = M_{Z_2}$ could
measure these $Z_2$ resonance effects.

\section*{\bf Vector Boson Pair Production}

In the SM, $W^+W^-$ and $WZ$ pair production can provide stringent tests of the
gauge theory since there are large cancellations between a $s$-channel gauge
boson amplitude and a $t$-channel fermion exchange amplitude. For example,
consider $p\bar p\to WZ + \rm anything$ production at $\sqrt s = 1.8$~TeV. The
components of the cross section are
\begin{equation}
\sigma(W^*) = 16\ {\rm pb}\,,\quad
\sigma(f) = 20\ {\rm pb}\,,\quad
\sigma(W^*,f) = -34\ {\rm pb}\,,\quad
\sigma_{\rm total}  = 2\rm\ pb\,,
\end{equation}
where $W^*$ denotes the $s$-channel resonance, $f$ the fermion exchange
contribution, and $(W^*,f)$ the interference contribution.
This cancellation is mandated by the asymptotic $s$-dependence of the cross
section\cite{collider}.
In the case of $WZ_2$ or $ZZ_2$ pair production the $s$-channel boson
contributions are highly suppressed by the mixing angle $\theta$. Consequently,
we expect much larger pair production cross sections than in the SM when
$M_{Z_2}$ is of order $M_Z$. Because the $Z'$ couplings to quarks are
universal, there is no reason for a cancellation of $s$- and $t$-channel
contributions\cite{cvet-lang-wz}.

The cross sections at the Tevatron
energy $\sqrt s = 1.8$~TeV are shown for $\lambda=1$
in Fig.~\ref{fig:sigma(ppbar)}.
We have included a $K$-factor of $K=1.3$ to approximate next-to-leading order
QCD contributions\cite{ohnemus}.
The cross sections for $WZ_2, Z_1 Z_2$, and
$\gamma Z_2$ scale linearly with $\lambda$.  For
comparison the corresponding SM cross sections for $WZ_1,\ \gamma Z_1$ and
$Z_1Z_1$ are indicated by squares on the figure.  We have imposed the
acceptance $p_T(\gamma)>15$ GeV and $|\eta(\gamma)|<1$ on the final state
photon.

The prospects for detecting the $Z_2$ would be best in the $Z_2\to b\bar b$
final state, with $b$-tagging by vertex detector or semileptonic decays to
reject backgrounds from light quarks and gluons in the $(\gamma,W,Z)jj$ final
state. The $b\bar b$ branching fraction of $Z_2$ is
\begin{equation}
B(Z_2\to b\bar b) = 0.2 \,.
\end{equation}
The signature of $WZ_2$ with $Z_2\to b\bar b$ are the same as those for Higgs
searches in $WH$ and $ZH$ final states with $H\to b\bar b$ decays.
Figure~\ref{fig:bbsigma} compares the $(W,Z,\gamma) b\bar b$ cross sections
for $\lambda=1$ with $(W,Z)H \to b\bar b$, where $B(H\to b\bar b)\approx 1$ for
the mass range shown.
We have included a $K$-factor of $K=1.25$ to approximate the next-to-leading
order QCD contributions\cite{han} to $WH$ and $ZH$ cross sections here.
 We see that for
$M_{Z_2}\approx105$~GeV the $W+(Z_2\to b\bar b)$ cross section with $\lambda=1$
is a factor 3 times the  $W+(H\to b\bar b)$ cross section.  Other
experimentally interesting
channels include $Z_1 + (Z_2\to b\bar b)$ and $\gamma + (Z_2\to b\bar b)$,
whose cross sections are also given in Fig.~\ref{fig:bbsigma}.
The $Z_1Z_2$ and $Z_2Z_2$ cross sections will give $b\bar b b\bar b,\ b\bar b
c\bar c,\ c\bar c c\bar c$ events above the SM QCD predictions.

The backgrounds to the $(\gamma,W,Z)Z_2$ signals in the $Z_2\to b\bar b$
channels arise from the $(\gamma,W,Z)g^*$ final states with a virtual
gluon $g^*$ giving a $b\bar b$ pair.
For the calculation of $p \bar p\to Wb\bar b$ we used the formulas in
Ref.~\cite{kunzst}, while we used MADGRAPH \cite{madgraph}
to calculate $p\bar p \to Zb\bar b$ and $p\bar p \to \gamma b\bar b$.
The differential cross sections $d\sigma/dm(b\bar b)$ for the signals and
backgrounds are shown in Figs.~\ref{fig:wbb}, \ref{fig:zbb}, and
\ref{fig:gbb} for $Wb\bar b$, $Zb\bar b$, and $\gamma b\bar b$ final
states, respectively.
While the background is a continuum in the $m(b\bar b)$ spectrum, the signal
gives a peak around the $Z_2$ mass.
In Figs.~\ref{fig:wbb}(a), \ref{fig:zbb}(a), and \ref{fig:gbb}, the solid
histogram is the SM background, while the long-dashed
and short-dashed histograms show the effect of the additional $Z_2$ boson
of mass 105 GeV and 85 GeV, respectively.
Figure \ref{fig:wbb}(b) and \ref{fig:zbb}(b) show the corresponding signals
and backgrounds for the SM Higgs boson of the same mass as the $Z_2$.
Since the $Z_2$ is universally coupled, the cross sections for $c\bar c$ final
states are the same as for $b\bar b$. If the UA2 bound of $\lambda\alt1$ were
weakened, larger $Z_2$ cross sections would be obtained with larger $\lambda$
values.

In Table~\ref{table:estimates} we present estimates of of the number of $Z_2\to
b\bar b$ signal and $b\bar b$ background events that would be expected at the
Tevatron in an invariant mass bin $\Delta m(b\bar b) = \pm10$~GeV  centered on
$M_{Z_2}=105$ GeV,
assuming 100~pb$^{-1}$ luminosity and 100\% detection efficiency.
The signal event rates are at the interesting level for $Z_2$ discovery.

\begin{table}[t]
\centering
\caption{\label{table:estimates}
Expected $Z_2\to b\bar b$ signals and background event rates at the Tevatron
with 100~pb$^{-1}$ luminosity and 100\% detection efficiency, for $\Delta
m(b\bar b) = \pm10$~GeV centered on $M_{Z_2}=105$ GeV.
A $Z_2$ coupling $\lambda=1$ is assumed. The event numbers in parentheses are
for acceptance cuts $p_T(b),p_T(\bar b)>10$~GeV, $|\eta(b)|,|\eta(\bar b)|<2$,
$|\cos\theta^*|<2/3$, where $\theta^*$ is the angle of the $b$ with respect to
the beam in the $b\bar b$ rest frame.}
\medskip
\tabcolsep=1.5em
\begin{tabular}{lcc}
\hline\hline
& Signal& Background\\
\hline
$s$-channel $Z_2$ & $6\times 10^4$ ($3\times10^4$)  &  $1.3\times10^7$
($6\times10^5$) \\
$WZ_2$ (with $W\to e\nu,\mu\nu$)& 9.6 & 6.1 \\
$Z_1Z_2$ (with $Z_1\to\nu\bar\nu$)& 3.3 & 4.9 \\
$Z_1Z_2$ (with $Z_1\to e\bar e, \mu\bar\mu$)&  1.1 &  1.7\\
$\gamma Z_2$  & 80 & 120 \\
\hline\hline
\end{tabular}
\end{table}

\section*{Summary}

In summary, we have shown the following:

\begin{itemize}

\item
A $Z'$ boson with baryonic couplings improves the overall fit to precision
electroweak observables, including LEP and SLD measurements along with other
low-energy measurements such as neutrino scattering.
Our conclusion in this regard is in agreement with other recent analyses which
were based on $Z$-pole observables only.

\item
The precision electroweak analysis constrains the product $\theta\lambda^{1/2}$
of the $Z,Z'$ mixing angle $\theta$ and the overall $Z'$ coupling strength
$\lambda^{1/2}$.

\item
The electroweak analysis favors a light $Z_2$ mass, $M_{Z_2}\alt 200$~GeV.
Values of $M_{Z_2}$ below the $Z$-mass are not ruled out.

\item
The $s$-channel production rate of $Z_2$ in $p\bar p$ collisions is constrained
by UA2 dijet measurements, with couplings up to $\lambda\sim1$ allowed.

\item
The $Z_2$ can be produced in association with $\gamma,W,Z$, with cross sections
at the
Tevatron exceeding corresponding cross sections for $Z_1$ production in
association with $\gamma,W,Z$.

\item
The $Z_2\to b\bar b$ decay mode is an important signal for $Z_2$ production in
association with $\gamma,W,Z$ at the Tevatron, giving a resonant enhancement in
the $b\bar b$ invariant mass spectrum above the QCD background. These processes
have a better signal-to-background ratio than the $s$-channel process.

\item
The $Z_2$ causes interference effects in $e^+e^-\to \bar bb(\bar cc)$ that
may be observable at LEP\,1.5. The interference contribution changes sign for $\sqrt s$ near $M_{Z_2}$ and is correlated with the signs of the deviations of $R_b(R_c)$ from SM predictions.

\end{itemize}

\section*{\bf Acknowledgments}

One of us (V.B.) thanks D.~Amidei, D.~Carlsmith,  T.~Han, J.~Huston,
W.-Y.~Keung, and P.~Mercadante for discussions. This research was initiated at
the Institute for Theoretical Physics at Santa Barbara, whose support is
gratefully acknowledged.
This research was supported in part by the U.S.~Department of Energy under
Grants No.~DE-FG02-95ER40896, No.~DE-FG03-93ER40757,
 and No.~DOE-EY-76-02-3071, in part by the National
Science Foundation Grant No. PHY94-07194, and in part
by the University of Wisconsin Research Committee with funds granted by the
Wisconsin Alumni Research Foundation.


\newpage
\section*{\bf Figures}

\begin{enumerate}

\item\label{fig:Z2_UA2}
The total cross section for the production of $p\bar p \to Z_2 \to jj$ at
$\sqrt{s}=630$ GeV for $\lambda=0.3$, 0.6, 1.  The UA2 90\% CL upper limit on
the production of a heavy boson decaying into 2 jets is shown.

\item\label{fig:Z2->bbar}
The $b\bar b$ invariant mass distribution and the inclusive $p_T(b)$
distribution at the Tevatron, including the contribution of a $Z_2$ resonance,
with $M_{Z_2}=105$~GeV and $\lambda=1$. The solid histogram denotes the SM
background, including the $Z_1\to b\bar b$ contribution. The dashed histogram includes the $Z_2$ contribution. The histograms at the bottoms of the figure are the $Z_2$ contributions alone. The acceptance cuts in
Table~\ref{table:estimates} are imposed.

\item\label{fig:resonance}
The $e^+e^-\to \bar bb(\bar cc)$ cross sections and the asymmetry $A_{\rm
FB}(b\bar b)$ versus $\sqrt{s}$ 
in the vicinity of a $Z_2$ resonance of mass $M_{Z_2}=105$~GeV,
with $\lambda=1$ and $\theta=-0.011$.
The solid curve denotes the SM background. The dashed and dot-dashed
curves include the contribution of an axial ($\gamma=0$) and vector
($\gamma=\pi/2$) baryonic $Z'$, respectively.  

\item\label{fig:sigma(ppbar)}
The total cross sections for the production of $p\bar p \to WZ_2, Z Z_2,
Z_2 Z_2, \gamma Z_2$ (solid curves) at $\sqrt{s}=1.8$ TeV.
The cross sections for the standard model $WZ, ZZ, \gamma Z$ (squares) and
$WH, ZH$ (dashed curves)  production are also shown.

\item\label{fig:bbsigma}
The  cross sections for the production of $p\bar p \to WZ_2, Z Z_2,
 \gamma Z_2$ followed by $Z_2 \to b\bar b$ (solid curves)
at $\sqrt{s}=1.8$ TeV.
The cross sections for the standard model $WZ, ZZ, \gamma Z$ with $Z\to b
\bar b$ (squares) and
$WH, ZH$ with $H\to b\bar b$ (dashed curves)  production are also shown.

\item\label{fig:wbb}
The $b\bar b$ invariant mass distribution $d\sigma/dm(b\bar b)$ for $Wb\bar b$
final state.  The solid histogram is the sum of the continuum $Wb\bar b$ and
the SM $Z$ boson, while the long-dashed and short-dashed histograms show the
additional $Z_2$ boson of mass 105 and 85 GeV, respectively.  Part~(b)
is similar to part~(a) with the additional $Z_2$ replaced by the
 SM Higgs boson.

\item\label{fig:zbb}
The $b\bar b$ invariant mass distribution $d\sigma/dm(b\bar b)$ for $Zb\bar b$
final state.  The solid histogram is the sum of the continuum $Zb\bar b$ and
the SM $Z$ boson, while the long-dashed and short-dashed histograms show the
additional $Z_2$ boson of mass 105 and 85 GeV, respectively.  Part~(b)
is similar to part~(a) with the additional $Z_2$ replaced by the
 SM Higgs boson.

\item\label{fig:gbb}
The $b\bar b$ invariant mass distribution $d\sigma/dm(b\bar b)$ for $\gamma
b\bar b$ final state.  The solid histogram is the sum of the continuum
$\gamma b\bar b$ and  the SM $Z$ boson, while the long-dashed and short-dashed
histograms show the additional $Z_2$ boson of mass 105 and 85 GeV,
respectively.


\end{enumerate}


\newpage
\pagestyle{empty}

\vspace*{9in}
\includegraphics{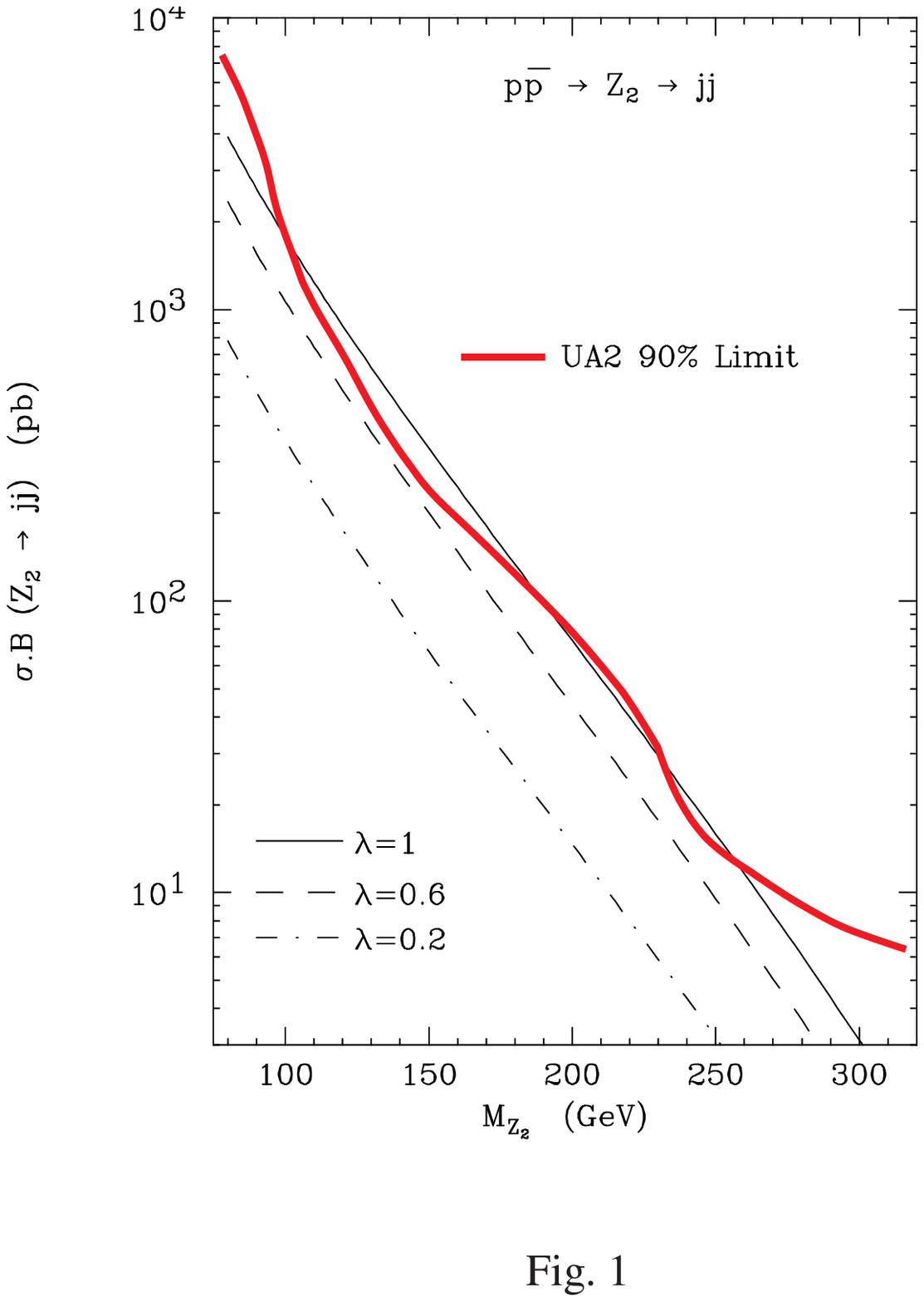}

\newpage
\vspace*{9in}
\includegraphics{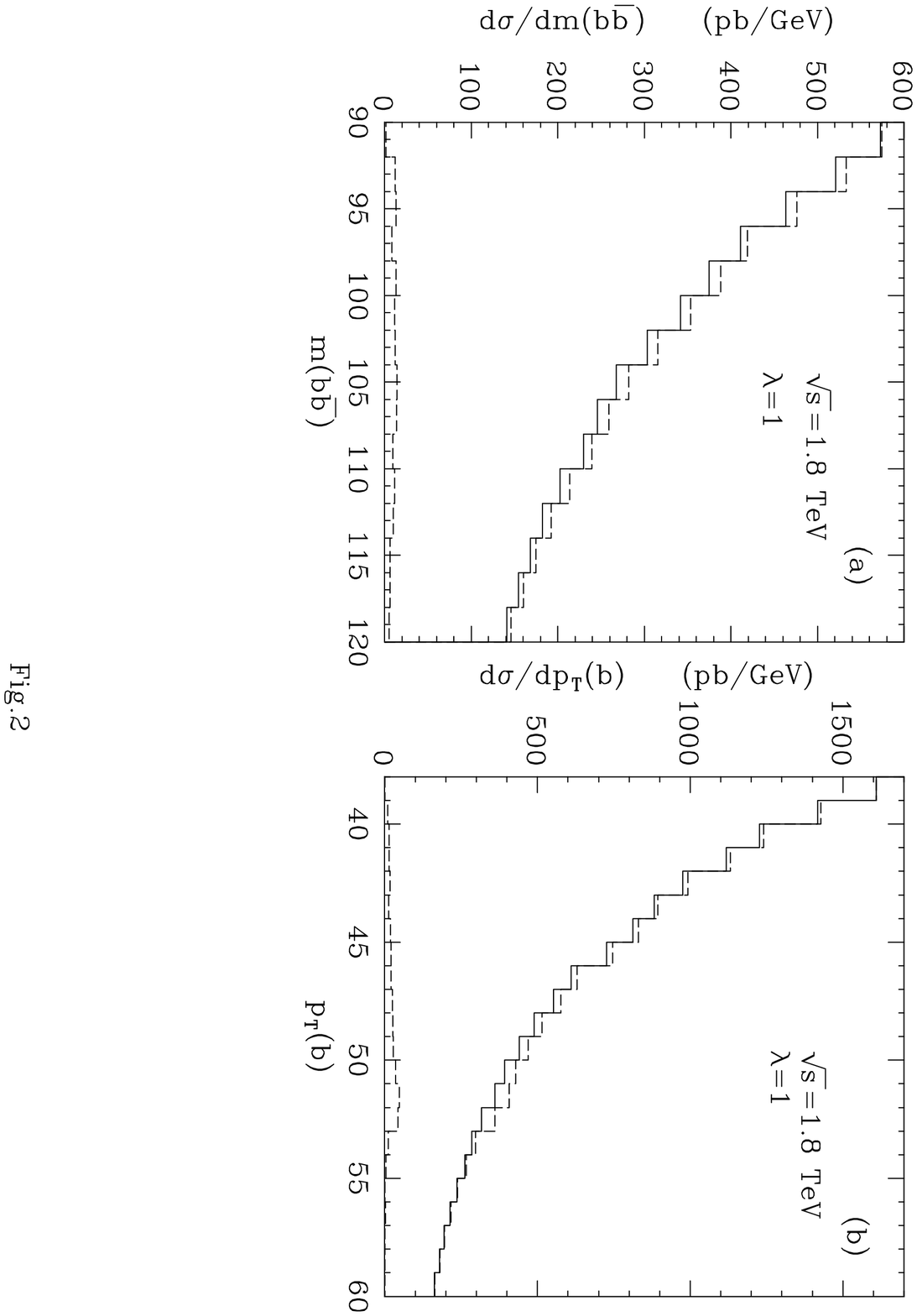}

\newpage
\vspace*{9in}
\includegraphics{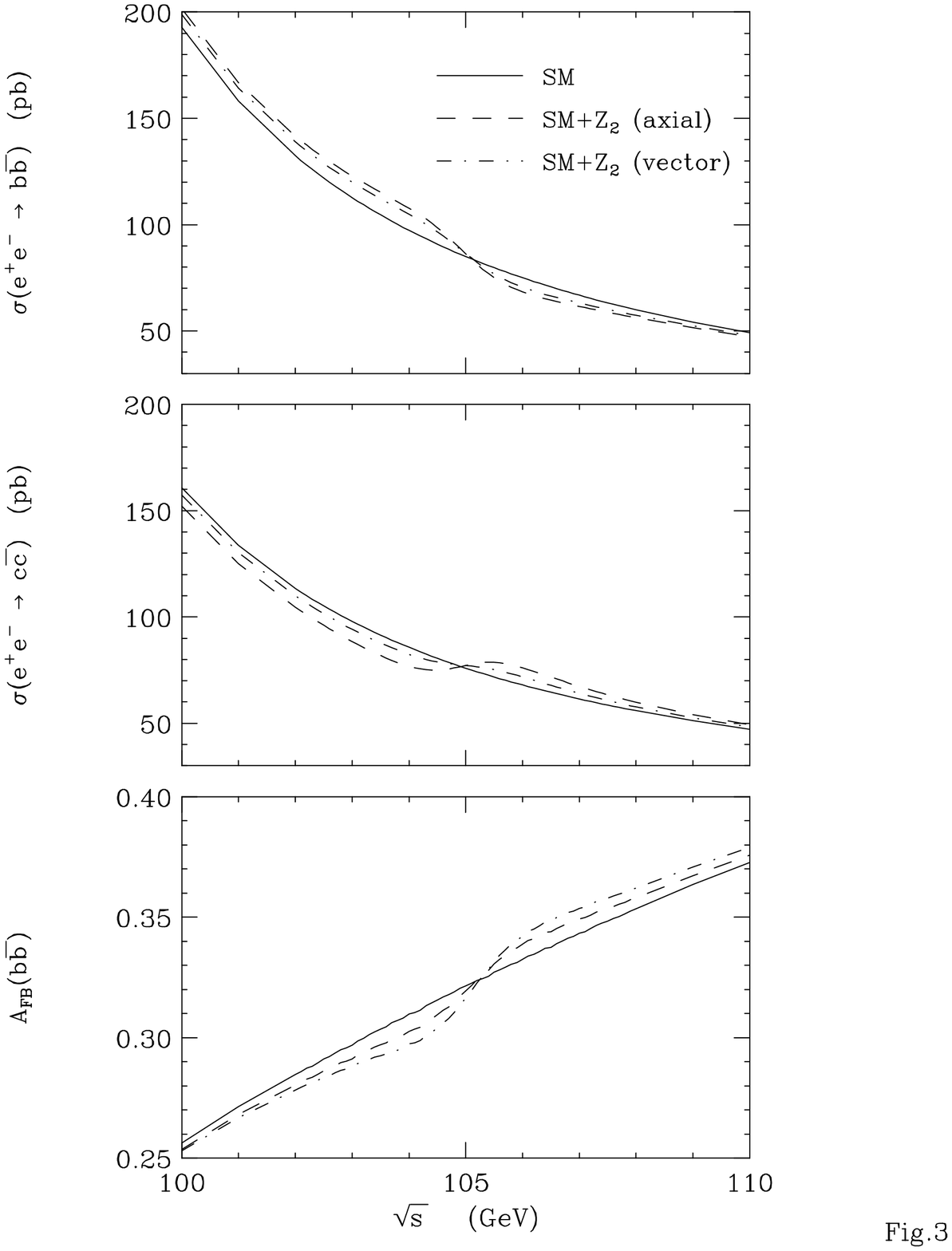}

\newpage
\vspace*{9in}
\includegraphics{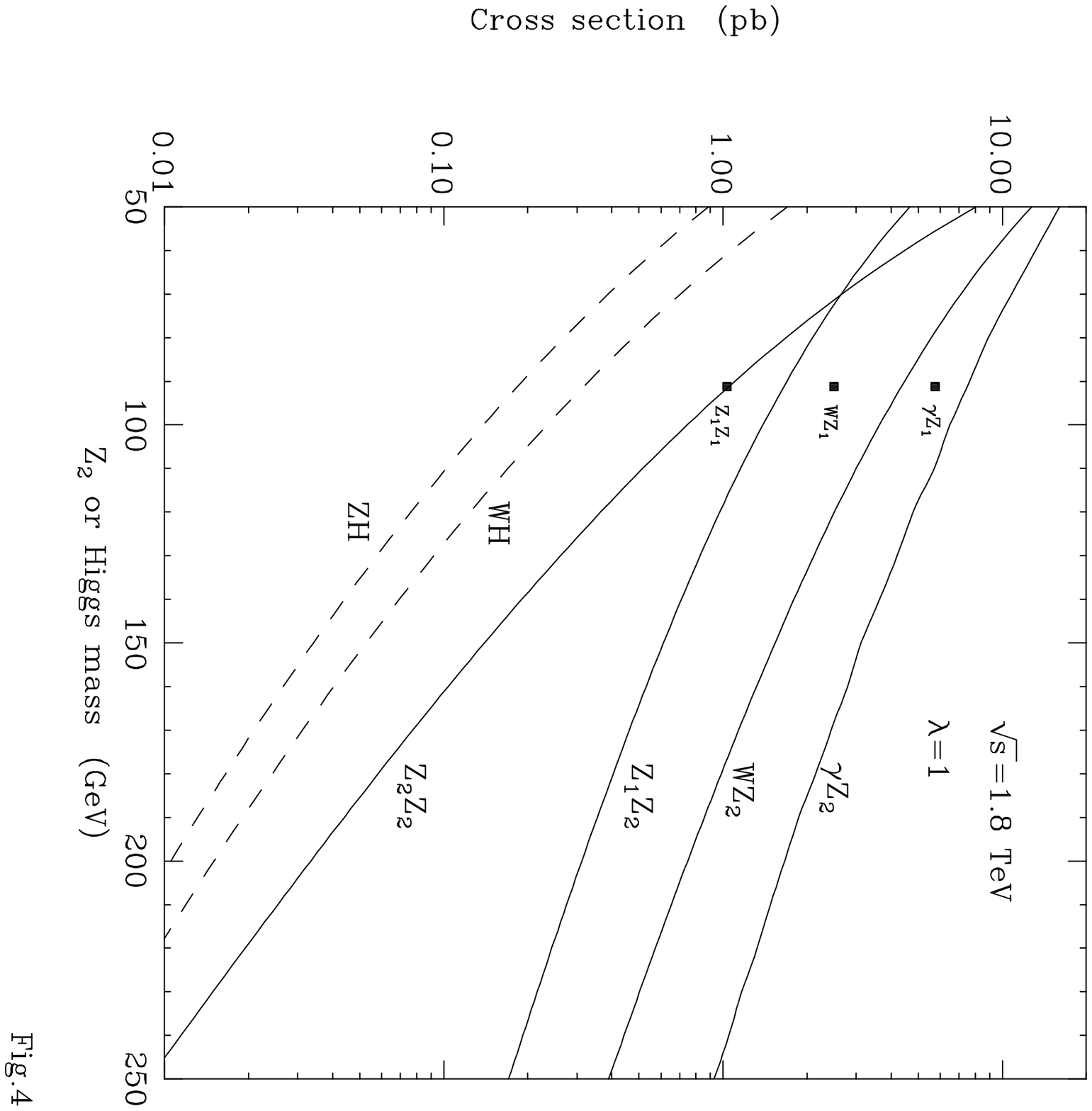}

\newpage
\vspace*{9in}
\includegraphics{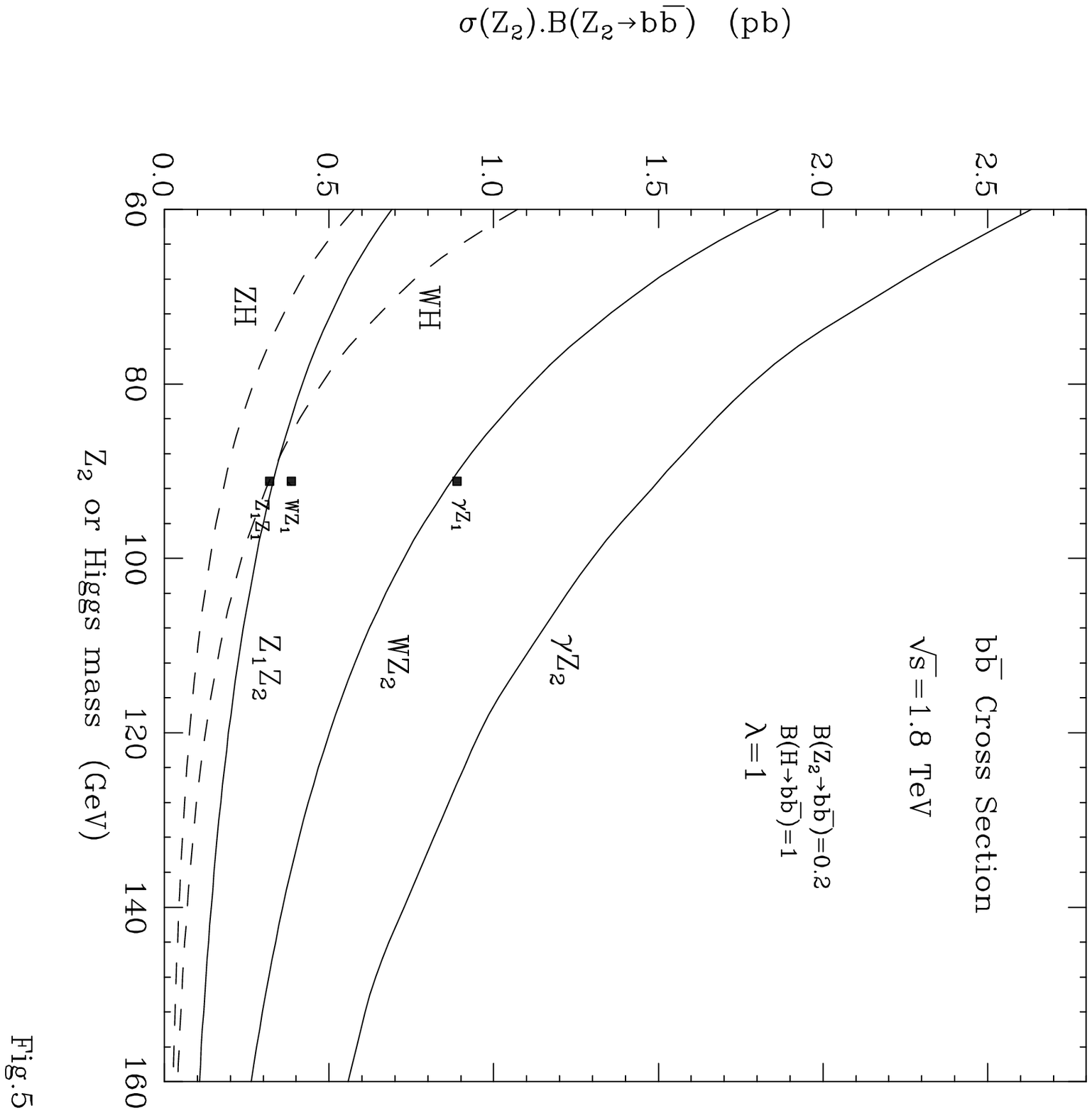}

\newpage
\vspace*{9in}
\includegraphics{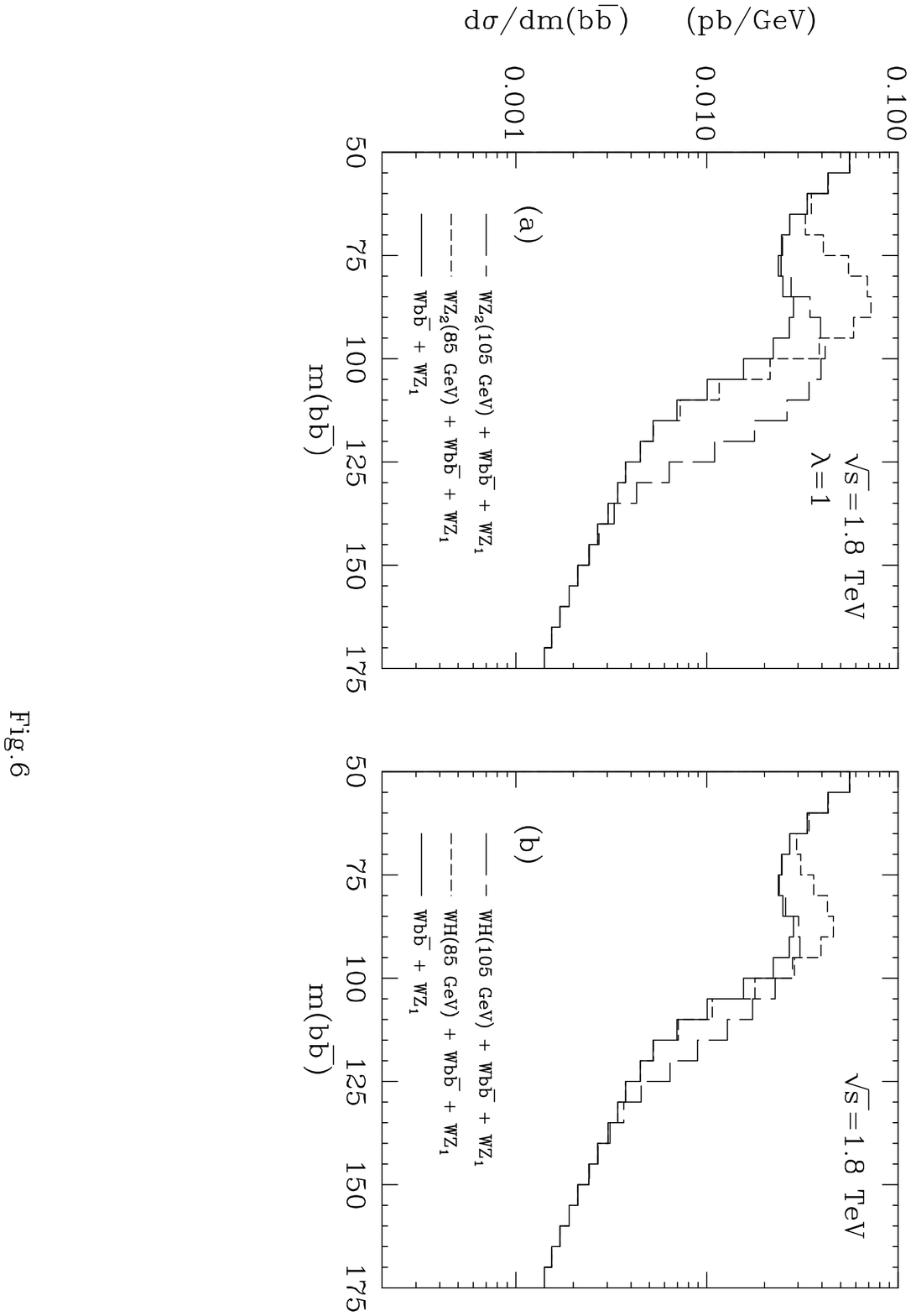}

\newpage
\vspace*{9in}
\includegraphics{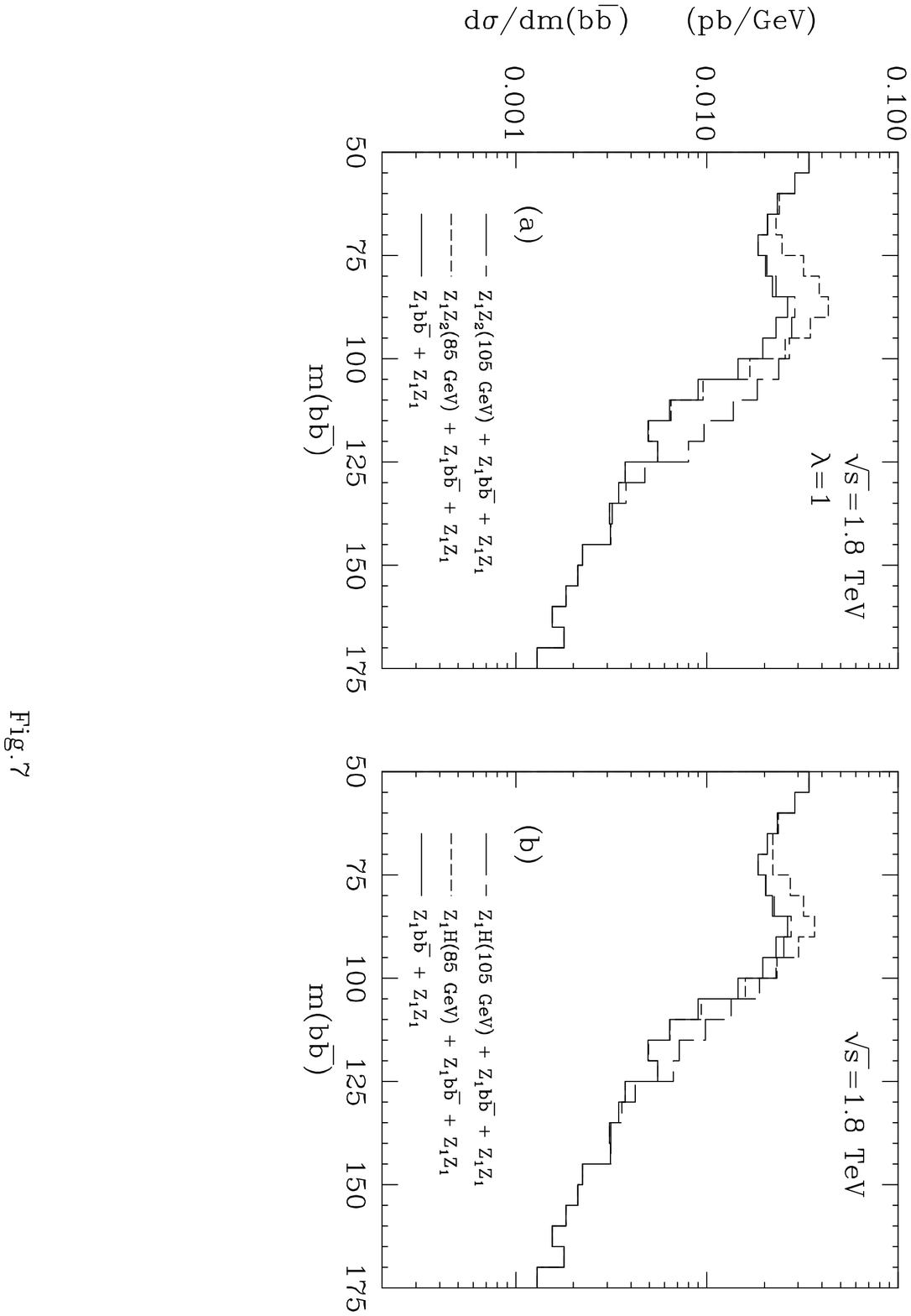}

\newpage
\vspace*{9in}
\includegraphics{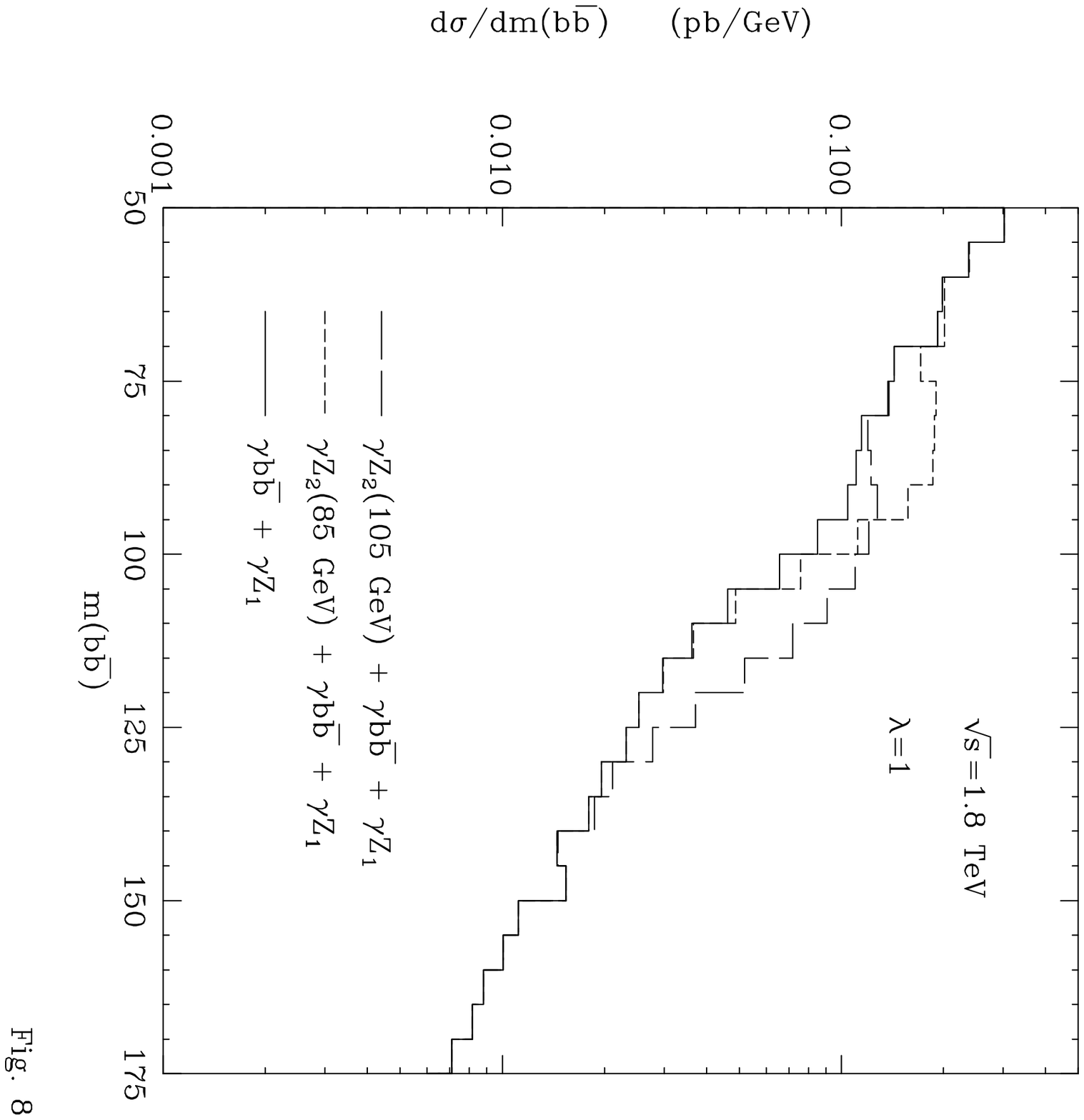}

\end{document}